\documentclass[aps,apl,twocolumn,groupedaddress,amsmath,reprint]{revtex4-1}
\usepackage[latin1]{inputenc}    % Accept european-encoded (latin1) characters.
\usepackage{graphicx}   % For eps figures
\usepackage{xspace}
\usepackage{color}

\begin{document}

\newcommand*{\DT}{\ensuremath{\Delta T}\xspace}
\newcommand*{\Te}{\ensuremath{T_{1/e}}\xspace}
\newcommand*{\Tqpt}{\ensuremath{T}_{\mathrm{1qp}}\xspace}
\newcommand*{\Ej}{\ensuremath{E}_{\mathrm{J}}\xspace}
\newcommand*{\Ec}{\ensuremath{E}_{\mathrm{C}}\xspace}
\newcommand*{\Tqp}{\ensuremath{\widetilde{T}_{\mathrm{1qp}}}\xspace}
\newcommand*{\Gqpt}{\ensuremath{\Gamma_{\mathrm{1qp}}}\xspace}
\newcommand*{\Gqp}{\ensuremath{\widetilde{\Gamma}_{\mathrm{1qp}}}\xspace}
\newcommand*{\Tr}{\ensuremath{T_{\mathrm{1R}}}\xspace}
\newcommand*{\xqp}{\ensuremath{x_\mathrm{qp} }\xspace}
\newcommand*{\nqp}{\ensuremath{n_\mathrm{qp} }\xspace}
\newcommand*{\nqpa}{\ensuremath{\langle n_\mathrm{qp} \rangle}\xspace}
\newcommand*{\Psw}{\ensuremath{P_\mathrm{sw}}\xspace}
\newcommand*{\tpw}{\ensuremath{t_\mathrm{pw}}\xspace}
\newcommand*{\rotq}{\ensuremath{\theta}\xspace}
\newcommand*{\Aq}{\ensuremath{A_\mathrm{Q}}\xspace}
\newcommand*{\Ai}{\ensuremath{A_\mathrm{I}}\xspace}
\newcommand*{\Hint}{\ensuremath{\hat{H}_\mathrm{int}}\xspace}
\newcommand*{\Hsub}{\ensuremath{\hat{H}_\mathrm{sub}}\xspace}
\newcommand*{\Hqb}{\ensuremath{\hat{H}_\mathrm{qb}}\xspace}
\newcommand*{\Htls}{\ensuremath{\hat{H}_\mathrm{TLS}}\xspace}
\newcommand*{\siX}{\ensuremath{\hat{\sigma}_\mathrm{x}}\xspace}
\newcommand*{\siY}{\ensuremath{\hat{\sigma}_\mathrm{y}}\xspace}
\newcommand*{\siZ}{\ensuremath{\hat{\sigma}_\mathrm{z}}\xspace}
\newcommand*{\siXq}{\ensuremath{\hat{\sigma}_\mathrm{x}^\mathrm{qb}}\xspace}
\newcommand*{\siYq}{\ensuremath{\hat{\sigma}_\mathrm{y}^\mathrm{qb}}\xspace}
\newcommand*{\siZq}{\ensuremath{\hat{\sigma}_\mathrm{z}^\mathrm{qb}}\xspace}
\newcommand*{\siXt}{\ensuremath{\hat{\sigma}_\mathrm{x}^\mathrm{TLS}}\xspace}
\newcommand*{\siYt}{\ensuremath{\hat{\sigma}_\mathrm{y}^\mathrm{TLS}}\xspace}
\newcommand*{\siZt}{\ensuremath{\hat{\sigma}_\mathrm{z}^\mathrm{TLS}}\xspace}
\newcommand*{\siXs}{\ensuremath{\hat{\sigma}_\mathrm{x}^\mathrm{sub}}\xspace}
\newcommand*{\siYs}{\ensuremath{\hat{\sigma}_\mathrm{y}^\mathrm{sub}}\xspace}
\newcommand*{\siZs}{\ensuremath{\hat{\sigma}_\mathrm{z}^\mathrm{sub}}\xspace}
\newcommand*{\Isq}{\ensuremath{I_\mathrm{b}}\xspace}
\newcommand*{\PhiQ}{\ensuremath{\Phi_\mathrm{qb}}\xspace}
\newcommand*{\fqb}{\ensuremath{\omega_\mathrm{qb}}\xspace}
\newcommand*{\ftls}{\ensuremath{f_\mathrm{TLS}}\xspace}
\newcommand*{\fosc}{\ensuremath{f_\mathrm{osc}}\xspace}
\newcommand*{\dph}{\ensuremath{\delta\Phi}\xspace}
\newcommand*{\df}{\ensuremath{\delta\!f}\xspace}
\newcommand*{\dpint}{\ensuremath{\delta\Phi_\mathrm{int}}\xspace}
\newcommand*{\TphN}{\ensuremath{T_{\varphi,\mathrm{N}}}\xspace}
\newcommand*{\Tph}[1]{\ensuremath{T_{\varphi,\mathrm{#1}}}\xspace}
\newcommand*{\tp}{\ensuremath{t_\mathrm{p}}\xspace}
\newcommand*{\fRabi}{\ensuremath{f_\mathrm{Rabi}}\xspace}
\newcommand*{\Ian}{\ensuremath{I_\mathrm{antenna}^\mathrm{mw}}\xspace}
\newcommand*{\Imr}{\ensuremath{I_\mathrm{r}^\mathrm{mw}}\xspace}
\newcommand*{\emd}{\ensuremath{\varepsilon^\mathrm{mw}_\mathrm{direct}}\xspace}
\newcommand*{\emt}{\ensuremath{\varepsilon^\mathrm{mw}}\xspace}

\newcommand{\ket}[1]{\vert  #1 \rangle}
\newcommand{\bra}[1]{\langle  #1 \vert}

\newcommand*{\PhiX}{\ensuremath{\Phi_\mathrm{X}}\xspace}
\newcommand*{\PhiZ}{\ensuremath{\Phi_\mathrm{Z}}\xspace}
\newcommand*{\fX}{\ensuremath{f_\mathrm{x}}\xspace}
\newcommand*{\fZ}{\ensuremath{f_\mathrm{z}}\xspace}
\newcommand*{\Ax}{\ensuremath{A_\mathrm{x}}\xspace}
\newcommand*{\Az}{\ensuremath{A_\mathrm{z}}\xspace}

\newcommand*{\TF}{\ensuremath{T_{\varphi F}}\xspace}
\newcommand*{\TE}{\ensuremath{T_{\varphi E}}\xspace}
\newcommand*{\GF}{\ensuremath{\Gamma_{\varphi F}}\xspace}
\newcommand*{\GE}{\ensuremath{\Gamma_{\varphi E}}\xspace}

\newcommand*{\GSs}{\ensuremath{\,\mathrm{GS/s}\xspace}}
\newcommand*{\mPh}{\ensuremath{\,\mathrm{m}\Phi_0}\xspace}
\newcommand*{\uPh}{\ensuremath{\,\mu\Phi_0}\xspace}

\newcommand*{\um}{\ensuremath{\,\mu\mathrm{m}}\xspace}
\newcommand*{\nm}{\ensuremath{\,\mathrm{nm}}\xspace}
\newcommand*{\mm}{\ensuremath{\,\mathrm{mm}}\xspace}
\newcommand*{\m}{\ensuremath{\,\mathrm{m}}\xspace}
\newcommand*{\sqm}{\ensuremath{\,\mathrm{m}^2}\xspace}
\newcommand*{\sqmm}{\ensuremath{\,\mathrm{mm}^2}\xspace}
\newcommand*{\squm}{\ensuremath{\,\mu\mathrm{m}^2}\xspace}
\newcommand*{\psqm}{\ensuremath{\,\mathrm{m}^{-2}}\xspace}
\newcommand*{\psqmV}{\ensuremath{\,\mathrm{m}^{-2}\mathrm{V}^{-1}}\xspace}
\newcommand*{\cm}{\ensuremath{\,\mathrm{cm}}\xspace}

\newcommand*{\nF}{\ensuremath{\,\mathrm{nF}}\xspace}
\newcommand*{\pF}{\ensuremath{\,\mathrm{pF}}\xspace}
\newcommand*{\pH}{\ensuremath{\,\mathrm{pH}}\xspace}

\newcommand*{\emob}{\ensuremath{\,\mathrm{m}^2/\mathrm{V}\mathrm{s}}\xspace}
\newcommand*{\edos}{\ensuremath{\,\mu\mathrm{C}/\mathrm{cm}^2}\xspace}
\newcommand*{\mbar}{\ensuremath{\,\mathrm{mbar}}\xspace}

\newcommand*{\A}{\ensuremath{\,\mathrm{A}}\xspace}
\newcommand*{\mA}{\ensuremath{\,\mathrm{mA}}\xspace}
\newcommand*{\nA}{\ensuremath{\,\mathrm{nA}}\xspace}
\newcommand*{\pA}{\ensuremath{\,\mathrm{pA}}\xspace}
\newcommand*{\fA}{\ensuremath{\,\mathrm{fA}}\xspace}
\newcommand*{\uA}{\ensuremath{\,\mu\mathrm{A}}\xspace}

\newcommand*{\Ohm}{\ensuremath{\,\Omega}\xspace}
\newcommand*{\kOhm}{\ensuremath{\,\mathrm{k}\Omega}\xspace}
\newcommand*{\MOhm}{\ensuremath{\,\mathrm{M}\Omega}\xspace}
\newcommand*{\GOhm}{\ensuremath{\,\mathrm{G}\Omega}\xspace}

\newcommand*{\Hz}{\ensuremath{\,\mathrm{Hz}}\xspace}
\newcommand*{\kHz}{\ensuremath{\,\mathrm{kHz}}\xspace}
\newcommand*{\MHz}{\ensuremath{\,\mathrm{MHz}}\xspace}
\newcommand*{\GHz}{\ensuremath{\,\mathrm{GHz}}\xspace}
\newcommand*{\THz}{\ensuremath{\,\mathrm{THz}}\xspace}

\newcommand*{\K}{\ensuremath{\,\mathrm{K}}\xspace}
\newcommand*{\mK}{\ensuremath{\,\mathrm{mK}}\xspace}

\newcommand*{\kV}{\ensuremath{\,\mathrm{kV}}\xspace}
\newcommand*{\V}{\ensuremath{\,\mathrm{V}}\xspace}
\newcommand*{\mV}{\ensuremath{\,\mathrm{mV}}\xspace}
\newcommand*{\uV}{\ensuremath{\,\mu\mathrm{V}}\xspace}
\newcommand*{\nV}{\ensuremath{\,\mathrm{nV}}\xspace}

\newcommand*{\eV}{\ensuremath{\,\mathrm{eV}}\xspace}
\newcommand*{\meV}{\ensuremath{\,\mathrm{meV}}\xspace}
\newcommand*{\ueV}{\ensuremath{\,\mu\mathrm{eV}}\xspace}

\newcommand*{\T}{\ensuremath{\,\mathrm{T}}\xspace}
\newcommand*{\mT}{\ensuremath{\,\mathrm{mT}}\xspace}
\newcommand*{\uT}{\ensuremath{\,\mu\mathrm{T}}\xspace}

\newcommand*{\ms}{\ensuremath{\,\mathrm{ms}}\xspace}
\newcommand*{\s}{\ensuremath{\,\mathrm{s}}\xspace}
\newcommand*{\us}{\ensuremath{\,\mathrm{\mu s}}\xspace}
\newcommand*{\ns}{\ensuremath{\,\mathrm{ns}}\xspace}
\newcommand*{\rpm}{\ensuremath{\,\mathrm{rpm}}\xspace}
\newcommand*{\minute}{\ensuremath{\,\mathrm{min}}\xspace}
\newcommand*{\degree}{\ensuremath{\,^\circ\mathrm{C}}\xspace}

\newcommand*{\EqRef}[1]{Eq.\,(\ref{#1})}
\newcommand*{\FigRef}[1]{Fig.\,\ref{#1}}
\newcommand*{\dd}[2]{\mathrm{\partial}#1/\mathrm{\partial}#2}
\newcommand*{\ddf}[2]{\frac{\mathrm{\partial}#1}{\mathrm{\partial}#2}}

\title{Suppressing relaxation in superconducting qubits by quasiparticle pumping}
 \author{Simon Gustavsson$^1$}
% \email{simongus@mit.edu}
 \author{Fei Yan$^{1}$}
 \author{Gianluigi Catelani$^{2}$}
 \author{Jonas Bylander$^{3}$}
 \author{Archana Kamal$^{1}$}
 \author{Jeffrey Birenbaum$^{4}$}
 \author{David Hover$^{4}$}
 \author{Danna Rosenberg$^{4}$}
 \author{Gabriel Samach$^{4}$}
 \author{Adam P. Sears$^{4}$}
 \author{Steven J. Weber$^{4}$}
 \author{Jonilyn L. Yoder$^{4}$}
 \author{John Clarke$^{5}$}
 \author{Andrew J. Kerman$^{4}$}
 \author{Fumiki Yoshihara$^6$}
 \author{Yasunobu Nakamura$^{7,8}$}
 \author{Terry P. Orlando$^{1}$}
 \author{William D. Oliver$^{1,4,9}$}

 \affiliation{$^1$Research Laboratory of Electronics, Massachusetts Institute of Technology, Cambridge, MA 02139, USA \\
  $^2$Forschungszentrum Jülich, Peter Grünberg Institut (PGI-2), 52425 Jülich, Germany\\
  $^3$Microtechnology and Nanoscience, Chalmers University of Technology, SE-41296 Gothenburg, Sweden\\
  $^4$MIT Lincoln Laboratory, 244 Wood Street, Lexington, MA 02420, USA\\
  $^5$Department of Physics, University of California, Berkeley, CA 94720\\
  $^6$The Institute of Physical and Chemical Research (RIKEN), Wako, Saitama 351-0198, Japan \\
  $^7$Center for Emergent Matter Science (CEMS), RIKEN, Wako, Saitama 351-0198, Japan\\
  $^8$Research Center for Advanced Science and Technology (RCAST), The University of Tokyo, Komaba, Meguro-ku, Tokyo 153-8904, Japan\\
  $^9$Department of Physics, Massachusetts Institute of Technology, Cambridge, MA 02139, USA}

%max 100 words abstract

\begin{abstract}
Dynamical error suppression techniques are commonly used to improve coherence in quantum
systems. They reduce dephasing errors by applying control pulses designed to reverse erroneous
coherent evolution driven by environmental noise. However, such methods cannot correct for irreversible processes such as energy relaxation. 
In this work, we investigate a complementary, stochastic approach to reducing errors:  instead of deterministically reversing the unwanted qubit evolution, we use control pulses to shape the noise environment dynamically.  In the context of superconducting qubits, we implement a pumping sequence to reduce the number of unpaired electrons (quasiparticles) in close proximity to the device.
We report a $70\%$ reduction in the quasiparticle density, resulting in a threefold enhancement in qubit relaxation times, and a comparable reduction in coherence variability. 
\end{abstract}

%%%%%%%%%%%%%%%%% END OF PREAMBLE %%%%%%%%%%%%%%%%

% Make the title.

\maketitle

% Introduction - Dynamical decoupling
Since Hahn's invention of the spin-echo in 1950 \cite{Hahn:1950}, coherent control techniques have been crucial tools for reducing errors, improving control fidelity, performing noise spectroscopy and generally extending coherence in both natural and artificial spin systems.
%  - improve dephasing
All of these methods are similar: they correct for dephasing errors by reversing unintended phase accumulations due to a noisy environment through the application of a sequence of control pulses, thereby improving the dephasing time $T_2$.
However, such coherent control techniques cannot correct for irreversible processes that reduce the relaxation time $T_1$, where energy is lost to the environment.
%  - instead, bath engineering
Improving $T_1$ requires either reducing the coupling between the spin system and its noisy environment, reducing the noise in the environment itself \cite{Murch:2013}, or implementing full quantum error correction. %\cite{Shor:1995}. %, Ofek:2016}.

We demonstrate a pumping sequence that dynamically reduces the noise in the environment and improves  $T_1$  of a superconducting qubit through an irreversible pumping process.
The sequence contains the same type of control pulses common to all dynamical-decoupling sequences, but instead of coherently and deterministically controlling the qubit time evolution, the sequence is designed to shape the noise stochastically via inelastic energy exchange with the environment. % in the environment via an irreversible pumping process involving inelastic energy exchange.
% previous work, cooling, optical pumping, 
Similar methods have been used to extend $T_2$ of spin qubits by dynamic nuclear polarization \cite{Reilly:2008}, and irreversible control techniques are commonly used to prepare systems into well-defined quantum states through optical pumping \cite{Happer:1972,Valenzuela:2006} and sideband cooling \cite{Wineland:1978}, % and to induce electromagnetic transparency \cite{Kocharovskaya:1986,Boller:1991}, 
but outside of quantum error correction, to our knowledge no dynamic enhancement of $T_1$ has been previously reported.

We implement the pumping sequence in a superconducting flux qubit, with the aim of reducing the population of unpaired electrons or quasiparticles in close vicinity to the device. 
% Introduction - quasiparticles
%In superconducting qubits, unpaired electrons or quasiparticles %(QPs) 
%are a major source of qubit decoherence.
%  - higher than expected density
As a superconducting circuit is cooled well below its critical temperature, the quasiparticle density via BCS theory is expected to be exponentially suppressed, but a number of experimental groups have reported higher-than-expected values in a wide variety of systems \cite{Aumentado:2004,DeVisser:2011,Maisi:2013,LevensonFalk:2014}.
%  - implications, qubits
Although the reasons for the enhanced quasiparticle population or the mechanism behind quasiparticle generation are not fully understood, their presence has a number of  adverse effects on the qubit performance, causing relaxation \cite{Lutchyn:2005,Martinis:2009,CatelaniPRL:2011,Leppakangas:2012,Sun:2012, Riste:2013}, dephasing \cite{Schreier:2008,Catelani:2012, DeGraaf:2013}, excess excited-state population \cite{Wenner:2013}, temporal variations in qubit parameters \cite{Pop:2014, Vool:2014, Bal:2015, Yan:2016}, and are predicted to be a major obstacle for realizing Majorana qubits in semiconductor nanowires \cite{Rainis:2012}. %Barends:2011,Corcoles:2011,
% our work - result
%When implementing the pumping sequence, we observe a $70\%$ reduction in the local quasiparticle population, leading to improved qubit relaxation times as well as reductions in temporal coherence variations.
%  - yale results - trapping
Our results provide an in situ technique for removing quasiparticles, especially in conjunction with recent experiments showing that vortices in superconducting electrodes can act as quasiparticles traps, thus keeping the quasiparticles away from the Josephson junctions where they may contribute to qubit relaxation \cite{Vool:2014, Wang:2014, Nsanzineza:2014,Taupin:2016}.

\begin{figure}[tb]
\centering
\includegraphics[width=1\linewidth]{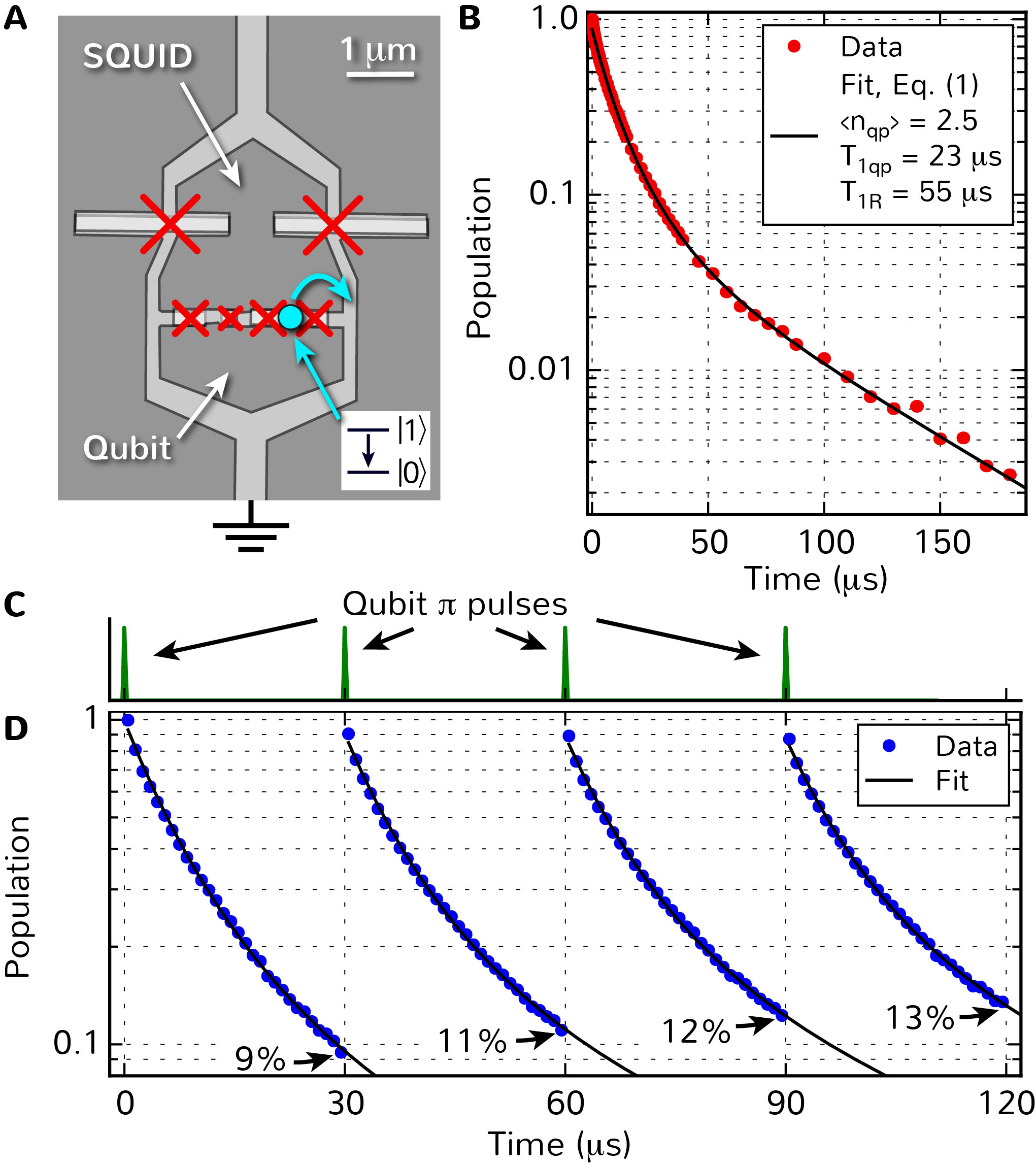}
\caption{Non-exponential decay in a superconducting flux qubit. (A) Schematic drawing of device A, consisting of a flux qubit (lower loop) coupled to a dc SQUID for qubit readout (outer loop).  The red crosses mark the position of the Josephson junctions.  Qubit relaxation is induced by quasiparticles tunneling across the qubit junctions, as illustrated by the turquoise circle. %The red/blue arrows illustrate the circulating currents that define the qubit basis states.
 (B) Qubit decay, measured by applying a $\pi$-pulse and delaying the qubit readout.  The decay is clearly non-exponential, with the solid line showing a fit to the decay function expected from quasiparticle tunneling [\EqRef{eq:decay} in the main text].
 (C) Pulse sequence for pumping quasiparticles away from the qubit junctions, consisting of multiple qubit $\pi$-pulses separated by a fixed period $\Delta T$. %(in this case $30\us$).
 (D) Average qubit population during the pumping sequence, measured by repeating the experiment over 40,000 trials.
% The first $\pi$ pulse puts the qubit in state $\ket{1}$.
 The remaining population after each pulse interval steadily increases, demonstrating that the qubit decay becomes progressively slower.
 The solid lines are fits to \EqRef{eq:decay} in the main text, with $\nqpa$ decreasing as $\{2.4, 1.9, 1.7, 1.6\}$ from the first to the last decay.
}
\label{fig1}
\end{figure}

% Non-exponential decay
We characterize and quantify the quasiparticle population by measuring qubit relaxation.
%  - decay from Fermi golden rule
Generally, the relaxation rate is given by a sum of contributions from many different decay channels. Quasiparticles contribute to the relaxation in a process whereby the qubit releases its energy to a quasiparticle tunneling across one of the Josephson junctions (\FigRef{fig1}A).
%  - describe non-exponential decay
Because of the small number of quasiparticles typically present in the device, fluctuations in the quasiparticle population lead to large temporal variations in the qubit decay rate. As a consequence, if the number of quasiparticles changes between trials while one repeats an experiment to determine the average qubit polarization, the time-domain decay no longer follows a single exponential, but rather takes the form \cite{Pop:2014} (see also Ref.~\cite{supp}, section S2)
\begin{equation}
p(t) =  e^{\nqpa \left( \exp{\left[-t/\Tqp\right]} -1 \right)} \, e^{-t/\Tr}.
 \label{eq:decay}
\end{equation}
Here, \nqpa is the average quasiparticle population in the qubit region during the experiment, \Tqp is the relaxation time induced by one quasiparticle and \Tr is the residual relaxation time from other decay channels such as flux noise, Purcell decay or dielectric losses. Because only quasiparticles are responsible for the non-exponential decay, \EqRef{eq:decay} provides a direct method for separating out quasiparticle contributions from other relaxation channels.

%  - two devices, new/old
The experiments are performed using two different flux qubits. Device A is a traditional flux qubit with switching-current readout using a dc superconducting quantum interference device (SQUID), while device B is a capacitively shunted (C-shunt) flux qubit \cite{Yan:2016}. Qubit A was operated at a frequency of $5.4\GHz$, while qubit B was operated at $4.7\GHz$, see sections S1 and S6 in Ref.~\cite{supp} for more information on qubit parameters.
%  - show fit, quantify n_qp and T1qp
Figure \ref{fig1}B shows the measured relaxation of qubit A, together with a fit to \EqRef{eq:decay}.  The decay is clearly non-exponential, exhibiting a  fast initial decay due to quasiparticle fluctuations, followed by a slower, constant decay due to residual relaxation channels. % for long delay times ($\tau\!>\!70\us$).
From the fit, we find an average quasiparticle population $\nqpa=2.5$, with $\Tqp=23\us$ and $\Tr=55\us$.
We have also measured the qubit decay as a function of flux and temperature to validate further its sensitivity to quasiparticles (sections S3 and S4 in Ref.~\cite{supp}).

% Pumping scheme - Fig 1
Interestingly, the same mechanism that leads to qubit relaxation also provides an opportunity for reducing the quasiparticle population. When the qubit relaxes through a quasiparticle tunneling event, the quasiparticle both tunnels to a different island and acquires an energy $\hbar \omega_0$ from the qubit ($\omega_0/2\pi$ is the qubit frequency). The increase in energy leads to a higher quasiparticle velocity (at constant mean free path), so that a quasiparticle can move more quickly 
%diffusion rate \note{Gianluigi: is there an expression for this, or a paper we can cite?}, allowing the quasiparticle to
away from the regions close to the qubit junctions where it may cause qubit relaxation. The situation is depicted in \FigRef{fig1}A, where the quasiparticle tunneling out from the section of the qubit loop containing the junctions may diffuse away towards the normal-metal ground electrode. %We can thus hope to reduce the quasiparticle population by purposefully letting the qubit relax through quasiparticle tunneling events.

%  - describe sequence and trace
We make use of this mechanism by applying a pulse sequence (\FigRef{fig1}C) consisting of several qubit $\pi$-pulses separated by a fixed period (in this case $\DT = 30 \us$).  The first $\pi$-pulse excites the qubit into state $\ket{1}$ and, during the subsequent waiting time, it has some probability of relaxing to the ground state.
%  - separation of time scales T_1qp < period < T_1R
Because  $\Tqp < \Tr$, this most likely occurs through a quasiparticle tunneling event, which transfers a quasiparticle across a junction, increases the quasiparticle energy and thereby enhances its diffusion rate.  The process is stochastic and may transfer quasiparticle in any direction, but by repeating the sequence we expect to pump quasiparticles away from the qubit junctions.
%  - extend diffusion time
The measured average qubit polarization during the pumping sequence (\FigRef{fig1}D) starts with the qubit in the ground state, the first $\pi$-pulse brings the qubit to $\ket{1}$, and during the following waiting period the qubit relaxes back to an average polarization of 9\%. The second $\pi$-pulse inverts the polarization to 91\%, and the qubit starts decaying again.  However, at the end of the second waiting period the remaining polarization is 11\%, demonstrating that the decay is slower during the second interval.  The third and fourth $\pi$-pulses further slow down the decay, yielding a remaining polarization of 12\% and 13\%, respectively. Note that the excess population can be removed by using single-shot readout techniques to reset the qubit state after the pumping sequence ends \cite{Geerlings:2013}.

\begin{figure}[tb]
\centering
\includegraphics[width=1\linewidth]{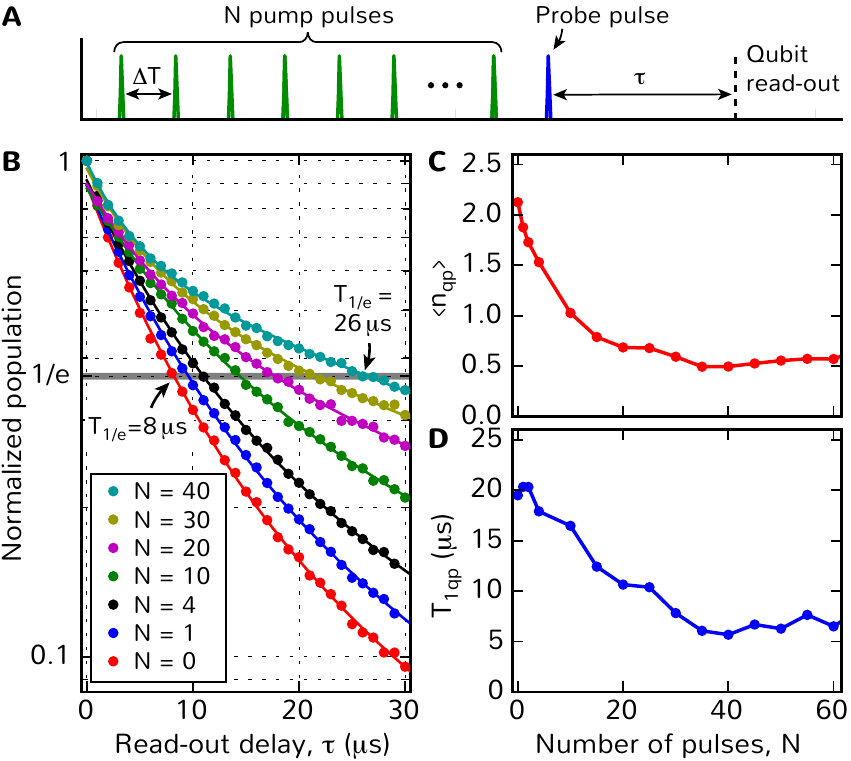}
\caption{Dynamic improvement in qubit decay time. (A) Pulse sequence for pumping quasiparticles. The last $\pi$-pulse acts as a probe pulse to measure the qubit polarization.
 (B) Normalized population vs. read-out delay, showing qubit decay after the pump sequence, measured with $\Delta T=10\us$ for increasing number of pulses $N$.  The decay time steadily increases from $\Te =8\us$ to $\Te=26\us$ after 40 pump pulses. The decay traces have been normalized to the population at $\tau=0$ to allow direct comparison. The solid lines are fits to \EqRef{eq:decay} in the main text. Each data point is averaged over $10^5$ trials.
 (C) Average quasiparticle number $\nqpa$ and (D) decay rate per quasiparticle $\Tqp$, extracted from the fits shown in panel B. $\Tr$ is held constant at $55\us$ for all fits.
}
\label{fig2}
\end{figure}

% Extend to many pulses - Fig 2
We quantify the reduction in qubit decay by extending the pump sequence to contain more $\pi$-pulses and fitting the decay to \EqRef{eq:decay}.
%  - data + fit
The measured qubit decay using up to 40 pumping pulses (\FigRef{fig2}) demonstrates a more than threefold enhancement in qubit decay time compared to the bare decay, where the decay time is defined as the time $\Te$ it takes for the signal to decay by a factor of $1/e$. The solid lines in \FigRef{fig2}B are fits to  \EqRef{eq:decay}, and Figs.\,\ref{fig2}C,D show the resulting fitting parameters $\nqpa$ and $\Tqp$ as a function of the number of pumping pulses.
%  - n_qp and T_1qp as a function of pulses
The average quasiparticle population drops from $\nqpa=2.2$ to about $\nqpa=0.5$ after 40 pulses, and then saturates at this level. At the same time, the decay time associated with one quasiparticle drops from $\Tqp=20\us$ to about $\Tqp \sim 7\us$.
%  - T_1qp change with n_qp?
The reduction of $\Tqp$ is somewhat unexpected, as one might generally expect the decay time per quasiparticle to remain constant as the quasiparticles are pumped away.  However, 
%at large $\nqpa$ there may be a number of excited quasiparticles present in the device, and these will 
as the number of $\pi$-pulses increases, the quasiparticles remaining near the junctions generally have higher energy and hence
cause qubit excitation as well as qubit relaxation;
since $1/\Tqp$ is the sum of decay and excitation rates, this conceptually explains in part the suppression of $\Tqp$. Note that, despite the introduction of an excitation rate, the qubit will still eventually decay to $\ket{0}$ due to non-quasiparticle relaxation channels ($\Tr$), preventing us from determining the excitation and decay rate separately from the steady-state qubit population.
We also note that the range of values of $\Tqp$ reported here is consistent with previous measurements in flux qubits \cite{Stern:2014}.  

To validate the quasiparticle pumping model further, and to rule out alternate explanations of the data, we have also implemented the same pumping scheme using pulses corresponding to $2\pi$ instead of $\pi$ rotations.  If the qubit's environment were directly influenced by the microwave pulses through a different mechanism than quasiparticle pumping (for example heating, or saturation of two-level systems), we would expect both the $\pi$ and $2\pi$ pulses to affect the qubit decay time. However, in the experiment we only observe an improvement in the qubit decay when driving the system with $\pi$ pulses, consistent with quasiparticle pumping model (Ref.~\cite{supp}, section S7).

% Return to equilibrium - Fig 3
Having demonstrated that the pumping sequence can substantially reduce the quasiparticle population, we investigate how long the reduction in $\nqpa$ persists before it returns to the equilibrium value by introducing a variable delay before the final probe pulse (see Ref.~\cite{supp}, section S5).
%  - rate for return, compare to Yale results
With the exception of an initial, faster rate for $t_\mathrm{delay}<50\us$, the return to its steady state is well described by an exponential function with a time constant of $300\us$. 
% justfiy T_1 << T_recovery
The timescale for quasiparticle recovery is much longer than the qubit lifetime, thus justifying the use of the steady-state solution in \EqRef{eq:decay} for estimating the quasiparticle population.

% New device - Fig 4
The measured quasiparticle population range of $\nqpa \sim 0.5 - 2.5$ corresponds to an upper bound on the normalized quasiparticles density of $\xqp\sim10^{-4} - 10^{-5}$, where $\xqp$ is the number of quasiparticles divided by the number of Cooper pairs and we assume that all decay-inducing quasiparticles are confined to the qubit islands.  %\note{Need to check the correct way to convert $n_{qp}$ to $x_{qp}$}.
This is higher than the typical values of $\xqp\sim10^{-6} - 10^{-7}$ reported in the literature \cite{Aumentado:2004,DeVisser:2011,Maisi:2013,LevensonFalk:2014}. %\cite{Aumentado:2004,Ferguson:2006,Shaw:2008, DeVisser:2011,Saira:2012,Maisi:2013,LevensonFalk:2014}. 
The difference may possibly be
attributed to the switching current readout of device A, where the qubit state is inferred by applying a short current pulse to the SQUID and determining its probability to switch to the normal state. Whenever a switching event occurs, quasiparticles are created in close vicinity to the SQUID junctions, leading to an increase in the overall quasiparticle density.

\begin{figure}[tb]
\centering
\includegraphics[width=1\linewidth]{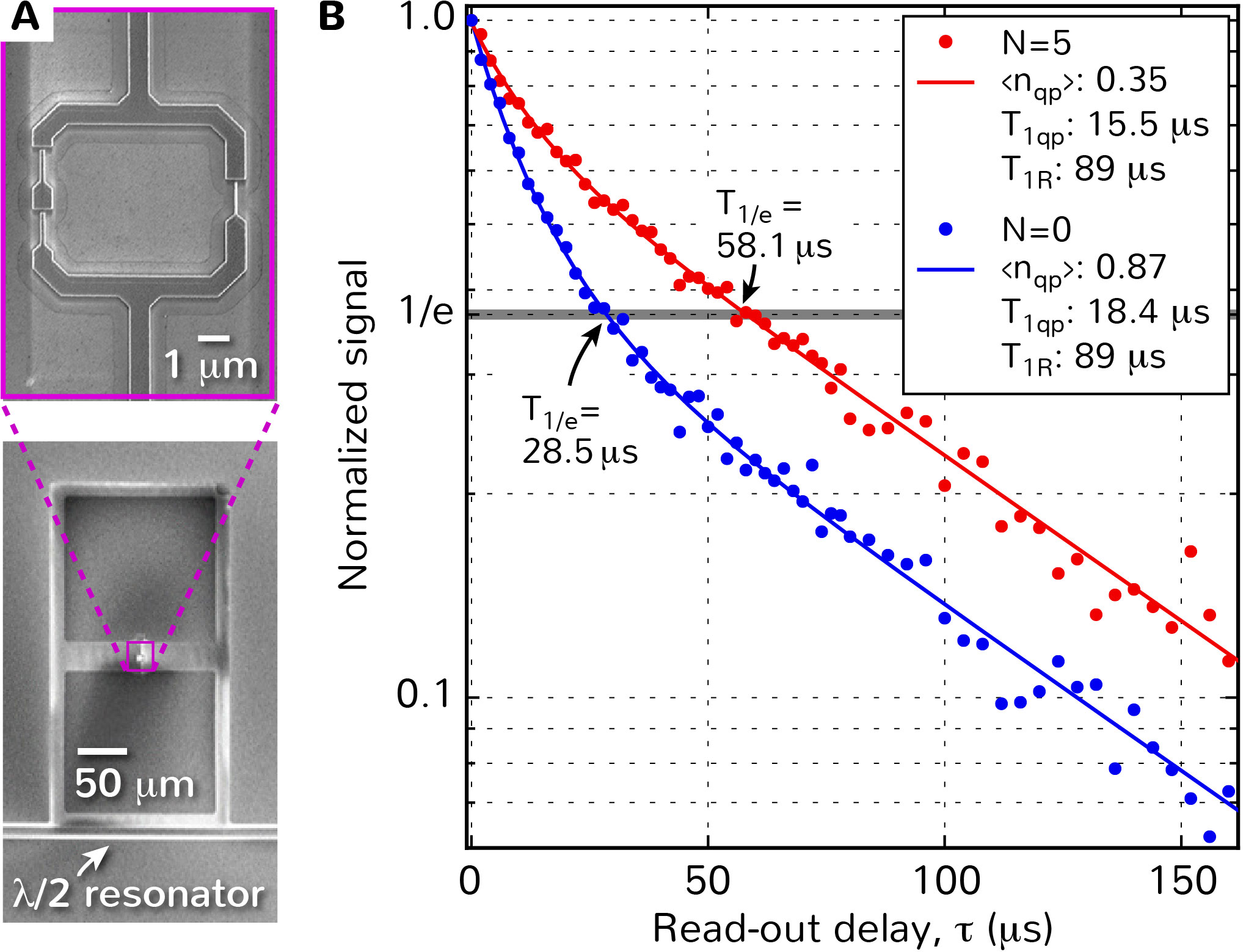}
\caption{Improvement in qubit decay time for a C-shunt flux qubit. (A) SEM image of device B, showing the large square capacitor plates (bottom panel) and a magnification of the qubit loop containing the three Josephson junctions (top panel).
 (B) Qubit relaxation, measured with and without quasiparticle pumping pulses.  The trace with $N=5$ pumping pulses was taken with a pulse period of $\DT=30\us$.   The data is averaged from 15 individual traces, acquired over a one-hour period.
 The fit was performed assuming the same value of $\Tr$ for both traces. The uncertainties in fitting parameters are $\nqpa$: $\pm0.02$, $\Tqp$: $\pm3.5\us$, and $\Tr$: $\pm4\us$.
}
\label{fig4}
\end{figure}

%  - describe device
We next investigate quasiparticle pumping in a dispersively read-out C-shunt flux qubit (device B), consisting of a flux qubit loop shunted by a large capacitance (\FigRef{fig4}A). Although the capacitor improves the qubit coherence by reducing its sensitivity to charge noise, the C-shunt flux qubit is still affected by quasiparticle fluctuations. As reported in Ref. \cite{Yan:2016}, the qubit was observed to switch between a stable configuration, with a purely exponential decay with $T_1>50\us$, and an unstable configuration with non-exponential decay and temporal fluctuations.  The switching between the various configurations was found to occur on a slow timescale, ranging from hours to several days. Similar switching events between stable and unstable configurations have also been observed in a fluxonium qubit, and were attributed to fluctuations in the quasiparticle density \cite{Vool:2014}.

We investigate how the quasiparticle pumping sequence affects the coherence of device B, both in stable and unstable configurations.
Since the switching between different configurations is random but slow, %we can average only over intervals during which no switching event has occurred.
we have taken care to average only over intervals when no switching event occurred.
%We have taken care to only average over intervals when no major switching event between different configurations occurred.
%  - works, 3x reduction in qp, 2x improvement in T1
Figure \ref{fig4}B shows the decay of device B, measured without and with $N=5$ quasiparticle pumping pulses.  The data were taken when the device was in a configuration where the qubit decay was clearly non-exponential, which is well captured by fits to \EqRef{eq:decay} (solid lines in \FigRef{fig4}B).  We observe a drop in the quasiparticle population from $0.87$ to $0.35$, leading to a two-fold enhancement in the qubit decay time. 
Note that the long-time decay rate is identical for both traces, as expected since the pumping scheme does not affect non-quasiparticle relaxation channels. 
The results demonstrate that the pumping scheme works even though device B does not have a ground electrode for trapping quasiparticles, but it has been shown that vortices in the capacitor pads can also act as quasiparticle traps \cite{Wang:2014}.
The pumping scheme should also be applicable to other qubit modalities where quasiparticle tunneling contributes to qubit relaxation.

\begin{figure}[tb]
\centering
\includegraphics[width=1\linewidth]{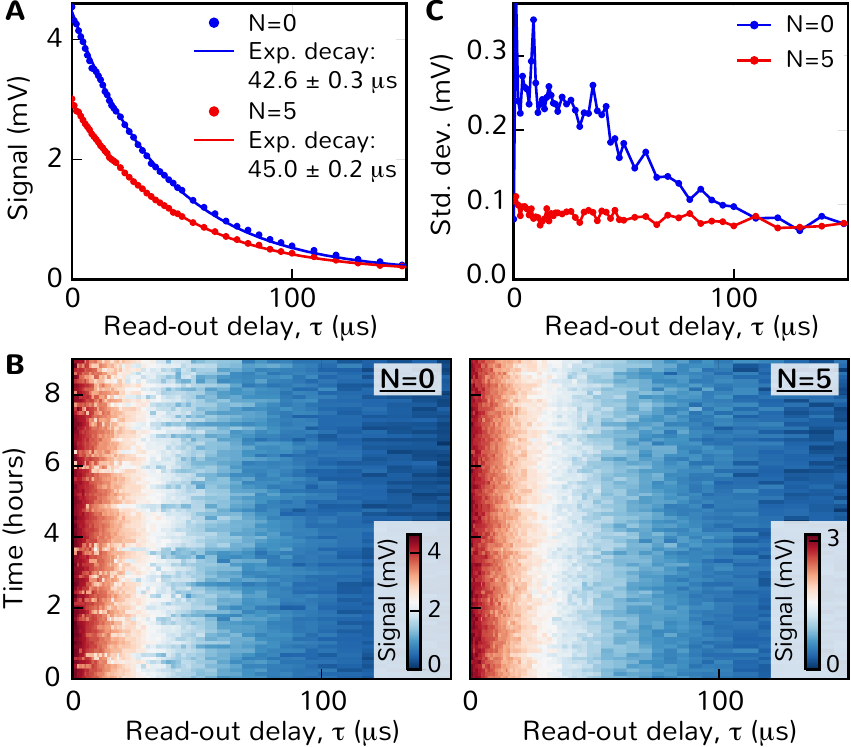}
\caption{Reduction of qubit coherence variations with quasiparticle pumping. (A) Averaged qubit decay, measured with and without pumping pulses. The decay function is close to exponential in both cases. The decay time increases by about $6\%$ with the pumping pulses. The traces have not been normalized to account for the decay during the pulse sequence, causing reduced contrast for the data with $N=5$. The data were measured with $\DT=30\us$.
 (B) Individual traces of the averaged decay data shown in panel A, measured without (left panel) and with 5 pumping pulses (right panel). The pumping sequence substantially reduces the temporal fluctuations observed in the decay without pumping pulses.
 (C) Standard deviation of the data in panel B, demonstrating the strong reduction in temporal shot-to-shot fluctuations in the presence of the pumping pulses.
 }
\label{fig5}
\end{figure}

% average over many traces
The data in \FigRef{fig4}B were acquired by continuously measuring qubit decay traces over a 1-hour period, and averaging them together.
% Reduction in fluctuations - Fig 5
Figure \ref{fig5} shows similar repeated measurements of the qubit decay with and without pumping pulses, but these traces were acquired about a week after the data in \FigRef{fig4}.  In the new data set, the qubit is in a configuration where the averaged decay function is relatively well described by a single exponential both with and without pumping pulses (\FigRef{fig5}A), and the five pumping pulses improve the decay time by only about $6\%$.
% traces
However, when investigating the individual decay traces (\FigRef{fig5}B), we find substantial amounts of noise and temporal fluctuations in the readout signal for the data without pumping pulses. These random variations vanish when  implementing the pumping sequence (right panel of \FigRef{fig5}B).

% std
To quantify the improvements in variability, we calculate the standard deviation of the read-out signal over 9 hours of data (\FigRef{fig5}C).
With pumping pulses, the standard deviation is independent of the read-out delay $\tau$, and can be ascribed to the noise of the HEMT amplifier used for amplification. Without pumping pulses, the standard deviation is substantially larger for $\tau<50\us$ but approaches the same level as for $N=5$ for long delay times.
%The increased noise at short $\tau$ is caused by fluctuations in the qubit population after the initial $\pi$ pulse, which go away as the qubit decays to the ground state.
The increased noise is caused by variations in the qubit $T_1$ time, which lead to strong fluctuations in the qubit population directly after the initial $\pi$-pulse. The fluctuations are reduced as the qubit decays to the ground states for long $\tau$, leaving only the contributions from the HEMT noise.

% conclusion
Our implementation of a stochastic scheme to shape the environment dynamically by pumping quasiparticles in a superconducting flux qubit leads to substantial improvements in both qubit coherence times and coherence variability.
In addition to applications in superconducting qubits, we anticipate the results to be of practical importance for implementing Majorana fermions in hybrid semiconductor/superconductor systems, where the presence of a single quasiparticle is detrimental to the device performance \cite{Rainis:2012}.

\begin{acknowledgments}
We gratefully acknowledge M. Blencowe, D. Campbell, M. Devoret, J. Grover, P. Krantz and I. Pop for useful discussions and P. Baldo, V. Bolkhovsky, G. Fitch, J. Miloshi, P. Murphy, B. Osadchy, K. Parrillo, R. Slattery, and T. Weir at MIT Lincoln Laboratory for technical assistance. This research was funded in part by the Office of the Director of National Intelligence (ODNI), Intelligence Advanced Research Projects Activity (IARPA) and by the Assistant Secretary of Defense for Research \& Engineering via MIT Lincoln Laboratory under Air Force Contract No. FA8721-05-C-0002; by the U.S. Army Research Office Grant No. W911NF-14-1-0682; and by the National Science Foundation Grant No. PHY-1415514. The views and conclusions contained herein are those of the authors and should not be interpreted as necessarily representing the official policies or endorsements, either expressed or implied, of ODNI, IARPA, or the US Government. GC acknowledges partial support by the EU under REA Grant Agreement No. CIG-618258. JB acknowledges partial support by the EU under REA Grant Agreement No. CIG-618353.
\end{acknowledgments}

\bibliographystyle{apsrev4-1}

\bibliography{Quasiparticle}

\renewcommand{\theequation}{S\arabic{equation}} 
\renewcommand{\thefigure}{S\arabic{figure}}
\renewcommand{\thetable}{S\arabic{table}}
\renewcommand{\thesection}{S\arabic{section}}
\setcounter{figure}{0}
\setcounter{equation}{0}

\clearpage

\setcounter{page}{1}

\clearpage
\begin{widetext}
\section*{Supplementary material}
%\tableofcontents

\section{Device A} \label{sec:devA}

Device A is a traditional flux qubit, with parameters and characteristics described in detail in Ref. \cite{Bylander:2011}. We operate the qubit with close to one-half flux quantum in the qubit loop, where the qubit frequency is $5.37\GHz$.  To resolve low-population features such as the long-time tail of the decay in Fig. 1B, we repeatedly measure $T_1$ traces and average them together.  The data in Fig. 1B are an average of 100 individual decay traces,  each  trace taking 10 minutes to acquire.  Each data point in an individual trace is averaged for 2000 trials, at a repetition period of $2\ms$. The slow repetition period is required to ensure the quasiparticles reach steady-state population between trials; if we run the experiments any faster, we observe modifications to the qubit decay function. This is consistent with the time scale needed for the quasiparticles to return to the steady-state population, as shown in \FigRef{fig3} in section \ref{sec:recovery}.

Between individual traces, we carefully measure the read-out signal that corresponds to the qubit ground state by averaging the read-out signal for $16\s$ without applying any pulses. This allows us to cancel out any drift in the measurement electronics that may occur during the long acquisition time, and is necessary for correct determination of the qubit polarization at low population.

\section{Non-exponential decay and population recovery} \label{sec:nonexp}

In this section we present a phenomenological model to support the validity of Eq.~(1) in the main text. Moreover, this model predicts an exponential recovery in $\nqpa$, in qualitative agreement with the data presented in \FigRef{fig3} in section \ref{sec:recovery}. We start by noting that
since quasiparticles must tunnel through one of the qubit junctions to cause relaxation, we can assume that only quasiparticles in a finite ``qubit region'' (between and close to the qubit junctions) should be taken into account, while material outside this region can be considered as a quasiparticle reservoir; although it is not clear what sets the size of the "qubit region", this open question does not affect the qualitative results of our phenomenological model.

We are interested in the probability $P(n,t)$ of having $n$ quasiparticles in the qubit region at time $t$. Quasiparticles can enter the qubit region from the reservoir at rate $\Gamma_{in}$ and leave with rate $\Gamma_{out}$. (The rate $\Gamma_{in}$ can also account for quasiparticle generation within the qubit region itself.) Then $P(n,t)$ obeys the rate equation
\begin{equation}\label{rate}
\dot{P}(n,t) = \Gamma_{in} P(n-1,t) - \Gamma_{in} P(n,t) - n \Gamma_{out} P(n,t) + (n+1) \Gamma_{out} P(n+1,t) \, ,
\end{equation}
where the dot denotes the time derivative.
The terms on the right hand side can all be motivated using simple physical arguments. For example, the second term accounts for the decrease of $P(n)$ due to a quasiparticle entering the qubit region and increasing $n\to n+1$, while the last term describes the increase in $P(n)$ due to one of $n+1$ quasiparticles leaving the region (since there are $n+1$ quasiparticles attempting to leave, the rate is multiplied by their number).

The steady-state solution $P_s$ of the above rate equation, obtained by setting $\dot{P}=0$, is a Poisson distribution
\begin{equation}\label{steady}
P_{s}(n) = \frac{\nqpa_s^n}{n!} e^{-\nqpa_s}, \qquad \nqpa_s = \frac{\Gamma_{in}}{\Gamma_{out}},
\end{equation}
where $\nqpa_s$ denotes the (steady-state) average quasiparticle number. As shown in Ref.~\cite{Pop:2014}, Eq.~(1) in the main text follows from the quasiparticle number having a Poisson distribution. We also note that, as explained in the main text, the $\pi$-pulse sequence increases the rate $\Gamma_{out}$ at which quasiparticles leave the qubit region, thus suppressing $\nqpa$.

We can use the relation $\nqpa \propto \Gamma_{out}^{-1}$ in Eq.~(\ref{steady}) to estimate a characteristic quasiparticle energy $\delta E$ (measured from the superconducting gap) before the pulses -- that is, the typical quasiparticle energy in the steady-state. Indeed, the rate $\Gamma_{out}$ is proportional to the quasiparticle group velocity (or equivalently, diffusion coefficient -- see Ref. ~\cite{Bardeen:1959}). The group velocity at energy $\epsilon$, in turn, is proportional to $\nu(\epsilon)^{-1}$, with $\nu(\epsilon)=\epsilon/\sqrt{\epsilon^2-\Delta^2}$  the normalized BCS density of states. We therefore conclude that $\nqpa \propto \nu(\epsilon)$, where $\epsilon$ denotes the characteristic quasiparticle energy. Let us indicate with $\nqpa_{N\pi}$ the average number of quasiparticles after $N$ pulses. Then, assuming $\delta E \ll \Delta$, we have $\nqpa_{0\pi} \propto \sqrt{\Delta/2\delta E}$. On the other hand, after a large number of pulses, $N\gg 1$, we can expect a quasiparticle still present in the qubit region to have a much larger energy, having absorbed energy $\omega$ from the qubit multiple times. If the acquired energy is at least comparable to the gap $\Delta$ itself, we can approximate $\nu(\epsilon) \approx 1$; we note that for this to be possible a necessary requirement is $N > \Delta/\omega \sim 11$. 

Empirically, in Fig.~2 both $\nqpa$ and $\Tqp$ saturate for $N \sim 35$, which is higher than the above lower limit -- this is to be expected, due to stochastic nature of the pumping process, and the possibility that during the pumping the quasiparticle loses energy via other processes, such as phonon emission (this process, however, is likely suppressed in small islands compared to the bulk, since the finite island size significantly alters the phonon spectrum). Therefore we conclude that for large pulse number $N$ we have $\nqpa_{0\pi}/\nqpa_{N\pi} = \sqrt{\Delta/2\delta E}$. Using this relation with $\Delta =0.233\,$meV (see the next section), $\nqpa_{0\pi}=2.2$, and $\nqpa_{N\pi}=0.5$ (see Fig. 2C), we find $\delta E\sim 70\,$mK, comparable to the estimated effective temperatures in this~\cite{Bylander:2011} and other~\cite{Pop:2014,Vool:2014,Riste:2013,Jin:2015} qubits.

From the above rate equation~(\ref{rate}) we can also derive the equation governing the time evolution of the average quasiparticle number $\nqpa(t)$. We multiply both sides by $n$ and sum over $n$ to find
\begin{equation}
\dot{\nqpa} = - \Gamma_{out} \nqpa + \Gamma_{in},
\end{equation}
which has the general solution
\begin{equation}
\nqpa(t) = \nqpa(0) e^{-\Gamma_{out} t} + \nqpa_s \left(1-e^{-\Gamma_{out} t}\right)\, .
\end{equation}
This expression shows that $\nqpa$ returns to its steady state over a time scale given by $1/\Gamma_{out}$. It also qualitatively explains the faster initial rise of $\nqpa$ in \FigRef{fig3}: immediately after the $\pi$-pulses, quasiparticles in the qubit regions have higher energy, so they leave at a faster rate; at later times, lower energy quasiparticles from the reservoir take the place of the higher-energy ones and $\Gamma_{out}$ returns to its lower, steady-state value.

Finally, from the recovery time $\Gamma_{out}^{-1} \sim 300\,\mu$s and the steady-state value $\nqpa_s \sim 2$ in \FigRef{fig3} of section \ref{sec:recovery}, we estimate $\Gamma_{in}^{-1} \sim 150\,\mu$s, see Eq.~(\ref{steady}). If we were to attribute this rate completely to generation of quasiparticles in the qubit region (as opposed to diffusion into it from the reservoir), we would obtain a generation rate $g\sim0.1\,$s$^{-1}$ for the normalized density $\xqp$. This rate is 2-3 orders of magnitude larger than the rates estimated in Refs.~\cite{Wang:2014,Vool:2014}. Thus, we conclude that the more likely mechanism determining $\Gamma_{in}$ is the arrival of quasiparticles generated outside the qubit, \textit{e.g.} from the readout SQUID junctions (cf. the main text).

\section{Temperature dependence} \label{sec:temp}

Figure \ref{fig:T1_vsTemp} shows the qubit decay time for device A as a function of the base temperature of the dilution refrigerator, where the measured decay traces have been fitted to a single exponential. While this fitting procedure is not justified at low temperatures due to
the clearly non-exponential decay caused by non-equilibrium
quasiparticles, we note that as temperature increases, the equilibrium (\textit{i.e.}, thermal) quasiparticle population increases, so that the quasiparticle decay channel dominates other relaxation mechanisms. Moreover, at large quasiparticle number the qubit decay becomes a simple exponential \cite{Pop:2014}, so fitting the data with an exponential is justified to study the effect of thermal quasiparticles.

The theoretical curve in Fig.~\ref{fig:T1_vsTemp} is calculated according to the procedures of Refs.~\cite{Paik:2011,Stern:2014}. We distinguish two contributions to $T_1$,
\begin{equation}\label{T1T}
\frac{1}{T_1} = \frac{1}{T_{1,ne}} + \frac{1}{T_{1,th}},
\end{equation}
where $T_{1,ne}$ accounts for all temperature-independent relaxation processes, such as dielectric losses (and possibly including non-equilibrium quasiparticles), while $T_{1,th}$ is due to thermal quasiparticles and is given by
\begin{equation}
\frac{1}{T_{1,th}} = \frac{16 E_J}{\pi\hbar} e^{-\Delta/k_BT}e^{\hbar \omega/2k_BT}K_0\left(\frac{\hbar\omega}{2k_BT}\right)\left(1+e^{-\hbar\omega/k_BT}\right)\left|\langle1|\sin\frac{\hat\varphi_L}{2}|0\rangle\right|^2.
\end{equation}
Here, $K_0$ is the modified Bessel function of the second kind, $\omega$ is the qubit frequency, $E_J=210\,$GHz is the Josephson energy of the large junctions, and $\varphi_L$ is the phase difference across one of the large junctions. We have taken into account that the thermal quasiparticle population in the small qubit islands is negligible.  The matrix element between ground and excited states in the last factor is calculated numerically using the known qubit parameters and we find:
\begin{equation}\label{lme}
\left|\langle1|\sin\frac{\hat\varphi_L}{2}|0\rangle\right| \simeq 0.240.
\end{equation}
In Eq.~(\ref{T1T}) there are two unknown parameters, $T_{1,ne}$ and the superconducting gap $\Delta$. We choose their values so that the theoretical curve bounds the experimental points from above, finding in particular $\Delta \simeq 0.233\,$meV.

% \note{Gianluigi, can you add an equation and say something about the fit here?}

\section{Flux dependence} \label{sec:flux}

Data points in Fig.~\ref{fig:T1_vsFlux} show the variation of $\Tqp$ with applied magnetic flux for device A, extracted by measuring the qubit decay as a function of magnetic flux and fitting the decay traces to Eq. (1) in the main paper. The lines in that figure show the expected theoretical dependence of $\Tqp$ on flux bias $f$, calculated as we now explain.

According to Ref.~\cite{CatelaniPRB:2011}, for low-energy quasiparticles each qubit junction $j$ contributes to the qubit relaxation rate by a term proportional to its Josephson energy $E_{Jj}$, the quasiparticle density $\xqp$ (averaged over the left and right sides of the junction), the density of states at the final quasiparticle energy $\sim \sqrt{\Delta/2\omega}$, and the matrix element between qubit states with the operator $\sin\hat\varphi_j/2$, where $\varphi_j$ is the phase difference across the junction.
Two factors can depend on the flux bias $f=\Phi/\Phi_0 - 1/2$, with $\Phi_0$ the flux quantum: the qubit frequency, and the matrix element. Indeed the qubit frequency has the usual form for flux qubits
\begin{equation}
\hbar \omega(f) = \sqrt{(\hbar \omega_0)^2+ (2I_p\Phi_0 f)^2}\, ,
\end{equation}
where $\omega_0$ is the frequency at zero flux bias and $I_p$ the persistent current. In the range of flux bias of interest the matrix elements are almost constant for the large junctions, taking the value given in Eq.~(\ref{lme}) (up to variations of less than one part in $10^{-3}$). On the other hand, the matrix element for the small junction (phase difference $\varphi_s$) is zero at $f=0$ due to an interference effect~\cite{Pop:2014}, and rises quickly with flux bias; as calculated numerically, it becomes comparable to the large junction matrix element at $f\approx 0.0019$. Collecting together the flux-independent quantities, we find
\begin{equation}\label{tqpf}
\frac{1}{\Tqp(f)} = \frac{1}{\Tqp(0)} \sqrt{\frac{\omega_0}{\omega(f)}}\left(1+\frac{\alpha}{\left|\langle1|\sin\frac{\hat\varphi_L}{2}|0\rangle\right|^2} \left|\langle1|\sin\frac{\hat\varphi_s}{2}|0\rangle\right|^2(f) \right),
\end{equation}
where $\alpha=0.54$ is the ratio between small and large junction Josephson energies. We show explicitly that the two flux-dependent factors are the qubit frequency and the small-junction matrix element. 
We note that for quasiparticles randomly distributed with equal probabilities among all qubit islands, the denominator in brackets would be larger by a factor of 3 \cite{CatelaniPRB:2011}; this would weaken the dependence of $\Tqp$ on flux, and worsen the comparison between theory and experiment. The smaller denominator accounts for the fact that that the central island is thicker than the two lateral ones, which means that its superconducting gap is smaller \cite{Yamamoto:2006}, and hence the quasiparticles are preferentially ``trapped`` there rather than being present in all the islands with equal probability. This preferential trapping might seem incompatible with the observation that even a single pulse decreases the qubit relaxation rate, see Figs. 1 and 2 in the main text. However, we remind that quasiparticles can tunnel without necessarily relaxing the qubit, via so-called parity switching events \cite{Riste:2013}. Therefore, if after the $\pi$-pulse the qubit relaxes by giving energy to a quasiparticle, the latter is now in a lateral island and from there it can tunnel either back to the central island or to the loop and then diffuse to ground; while the gap difference hinders the parity-switching events when the quasiparticle is in the central island, no such energy barrier is present for a quasiparticle in a lateral island. Adding more $\pi$ pulses facilitates the further tunneling of the quasiparticles between islands; perhaps it might be possible to model the quasiparticle dynamics as a random walk between the islands with absorbing boundary conditions due to ground,  but this lies beyond the scope of the present work.

In Eq.~(\ref{tqpf}) the only unknown parameter is $\Tqp(0)$; however, in extracting $\Tqp(f)$ from the qubit decay measured at different flux biases by fitting with Eq.~(1) in the main text, we have also the freedom to choose $\nqpa$ (in that equation we set $\Tr=55\,\mu$s , as obtained independently -- cf. Fig.~1 in the main text). We take $\nqpa=1.5$ to obtain $\Tqp(0)=23~\mu$s (see Fig.~1 in the main text) as the optimal choice to fit the $\Tqp(f)$ data points extracted from the decay measurements. Different choices of $\nqpa$ would lead to different optimal values for $\Tqp(0)$; the dashed lines in Fig.~\ref{fig:T1_vsFlux} are obtained for $\nqpa=1.8$ and 1.2 (upper and lower line, respectively). Almost all the points fall between these two lines, indicating that fluctuations in $\nqpa$ are the likely cause of noise in the experimental data. In fact, fluctuations in $\nqpa$ have been observed in a fluxonium qubit~\cite{Pop:2014}.

\section{Recovery of quasiparticle population} \label{sec:recovery}

%  - describe experiments
Figure \ref{fig3} shows the result of an experiment where 20 pumping pulses were used to reduce $\nqpa$ to 0.6, but where a variable delay was introduced before the final probe pulse (see inset of \FigRef{fig3}).
%  - rate for return, compare to Yale results
We find that it takes more than $1\ms$ for the quasiparticle population to return to steady-state. With the exception of an initial, faster rate for $t_\mathrm{delay}<50\us$, the return to its steady state is well described by an exponential function with a time constant of $300\us$. In our experiments, it is not possible to establish what determines the recovery time (it is related to the rates at which quasiparticles leave and re-enter the qubit, see Ref.~\cite{supp}, section S2),
%is due to quasiparticle generation and/or diffusion from other parts of the device \cite{Wang:2014},
but we note that the time needed to reach steady state is comparable to timescales reported for quasiparticles in a fluxonium qubit \cite{Vool:2014}.
% justfiy T_1 << T_recovery
We also note that timescale for quasiparticle recovery is much longer than the qubit lifetime, thus justifying the use of the steady-state solution in \EqRef{eq:decay} for estimating the quasiparticle population.

\section{Device B} \label{sec:devB}

Device B is a C-shunt flux qubit, described in detail in Ref. \cite{Yan:2016} along with the experimental setup. We operate the qubit at flux bias $f=0$, where the qubit frequency is $4.7\GHz$.

Since this device showed switching between configurations with different decay behavior on very slow time scales (hours to days), it became particularly important to ensure that the qubit was in the same configuration when comparing decay traces with and without pump pulses. For the data in Figs. 3 and 4 in the main paper, we used the following procedure to compare the qubit decay with $N=0$ and $N=5$ pumping pulses.
\begin{itemize}
\item For each value of read-out delay, $\tau$, from 0 to $160\us$:
\begin{itemize}
\item Determine qubit population without pumping pulses by averaging over 1000-2000 trials.
\item Determine qubit population with $N=5$ pumping pulses by averaging over 1000-2000 trials.
\end{itemize}
\item Repeat the above procedure 20-100 times to improve statistics.
\end{itemize}
By switching between $N=0$ and $N=5$ before going on to the next value of $\tau$, we ensure that any slow drift and switching behavior occurring on long timescales does not affect the decay traces for $N=0$ and $N=5$ differently.

\section{Pumping with $2\pi$ pulses} \label{sec:2pi}

To further validate the quasiparticle pumping scheme, and to rule out that the microwave pulses for qubit driving interact directly with the qubit's environment, we have also implemented the pumping scheme using pulses corresponding to $2\pi$ instead of $\pi$ qubit rotations.  If the qubit's environment were directly influenced by the pumping pulses through a different mechanism than qubit-quasiparticle interactions, we would expect both the $\pi$ and $2\pi$ pulses to affect the qubit decay time.  However, we would not expect the $2\pi$ pulses to work for quasiparticle pumping, since each $2\pi$ pulse will leave the qubit state unchanged.

The experiments were conducted using Device C, which is a C-shunt flux qubit with the same layout and circuit design as Device B, but fabricated in a different fabrication run with a slightly different value of the critical current density, giving a qubit frequency of $3.7\GHz$ (when biased at half-flux quantum).  Figure~\ref{fig:2pi} shows the results of experiments similar to the ones of Figure 4 in the main paper, but with the addition of implementing the pumping sequence with pulses that correspond to both $\pi$ and $2\pi$ qubit rotations.
Fig.~\ref{fig:2pi}A shows decay traces of the qubit measured both without and with $\pi$ as well as $2\pi$ pumping pulses, averaged for 45 hours. Device C shows a cleaner exponential devices than both devices A and B, but there is still a $3\%$ improvement in the decay time when applying pumping pulses that correspond to qubit $\pi$ rotations.  However, there is no improvement when using $2\pi$ rotations.  

Figure~\ref{fig:2pi}B shows the individual traces of the averaged curves plotted in Fig.~\ref{fig:2pi}A.  Similarly to Fig. 4 in the main paper, the temporal fluctuations are visibly reduced when applying $\pi$ pumping pulses compared to without or when pumping with $2\pi$ pulses.
To quantify the improvements in variability, we calculate the standard deviation of the read-out signal over the 45 hours of data (\FigRef{fig:2pi}C).  When using $2\pi$ pulses, the variations are identical to the variations measured without pumping pulses.  However, the variations are substantially suppressed with $\pi$ pulses.

The results of Fig.~\ref{fig:2pi} clearly shows that the quasiparticle pumping scheme only works when driving the qubit with $\pi$ pulses.

\clearpage

\begin{figure}[t]
\centering
\includegraphics[width=0.6\linewidth]{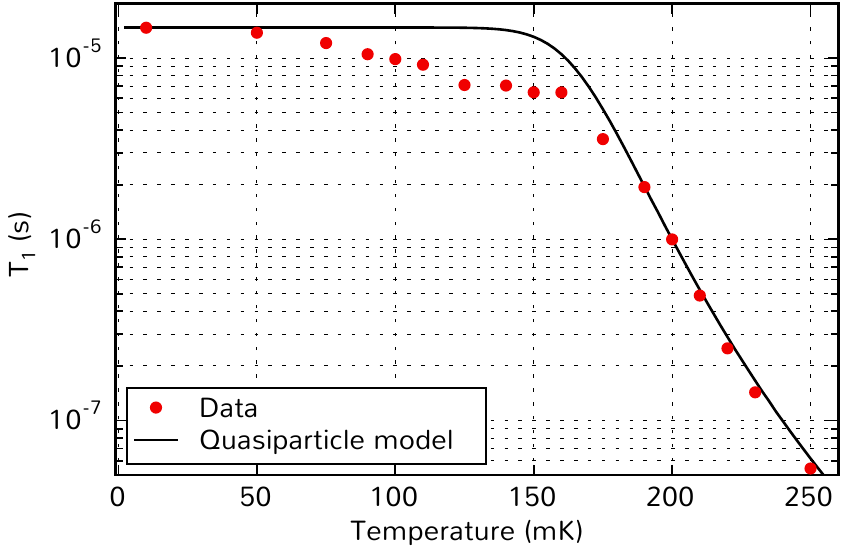}
\caption{Qubit relaxation time as a function of refrigerator temperature. The relaxation times are extracted by fitting decay traces to a single exponential. The black line is the decay time expected from combining the effects of temperature-independent relaxation channels and of thermal quasiparticles -- see text for details.
%, assuming that quasiparticles is the only source of relaxation at low temperatures.
}
\label{fig:T1_vsTemp}
\end{figure}

\clearpage

\begin{figure}[t]
\centering
\includegraphics[width=0.6\linewidth]{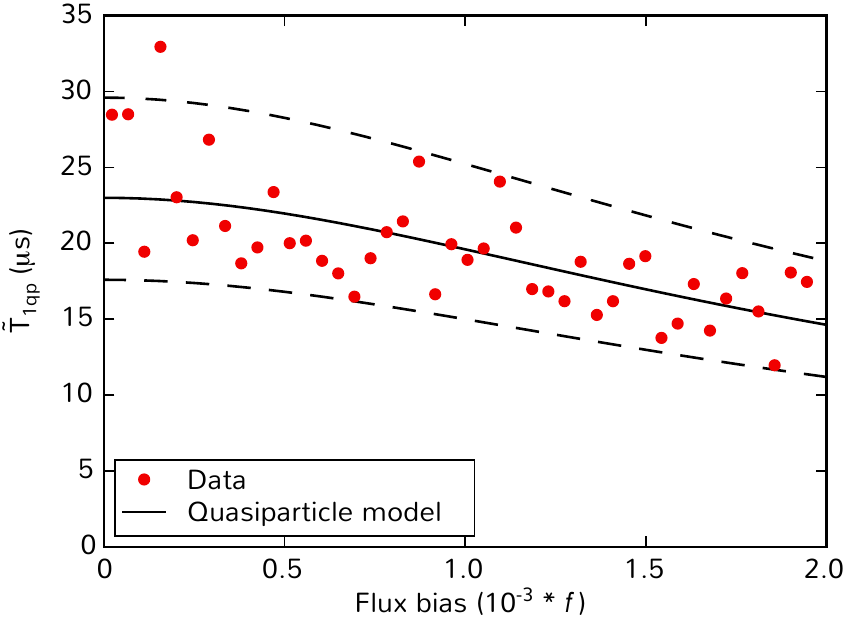}
\caption{Qubit decay time as a function of magnetic flux bias. The single-quasiparticle induced qubit relaxation time $\Tqp$ is measured as a function of applied magnetic flux bias $f=\Phi/\Phi_0 - 1/2$. The experimental values of $\Tqp(f)$ are obtained by fitting individual decay traces to Eq. (1) in the main text with
$\nqpa=1.5$ and $\Tr=55\us$ held fixed for all flux values.
The solid line shows the expected behavior of $\Tqp(f)$ -- see text for details of the calculation. The dashed lines correspond to a $\pm0.3$ change in $\nqpa$, showing that fluctuations in $\nqpa$ can account for the noise in the data.
}
\label{fig:T1_vsFlux}
\end{figure}

\clearpage

\begin{figure}[t]
\centering
\includegraphics[width=0.6\linewidth]{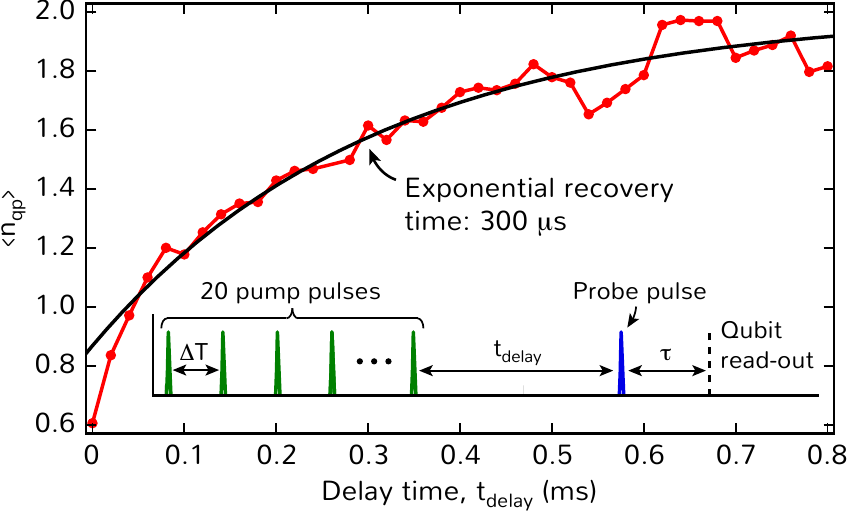}
\caption{Recovery of quasiparticle population $\nqpa$ after the pump sequence.  The population initially rises to about $1$ during the first $50\us$, then drifts back towards the equilibrium value on a slower time scale.
The data are measured by applying 20 pump pulses separated by $\Delta T=10\us$, and delaying the final probe pulse that measures the qubit decay function.  The average population $\nqpa$ is extracted by fitting the qubit decay after the probe pulse to \EqRef{eq:decay} in the main text.
 }
\label{fig3}
\end{figure}

\clearpage

\begin{figure}[t]
\centering
\includegraphics[width=0.7\linewidth]{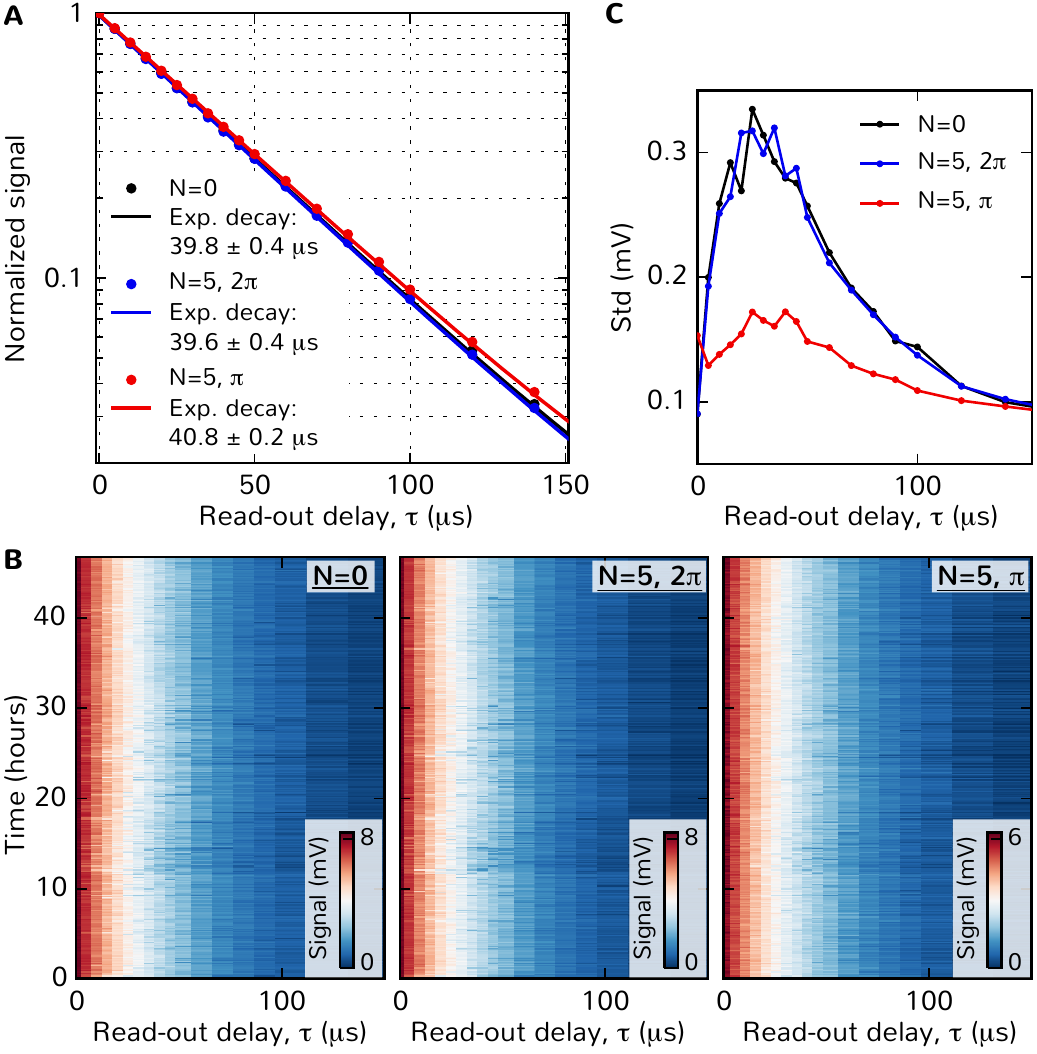}

\caption{Pumping with $2\pi$ pulses. (A) Averaged qubit decay for device C, measured without and with either five $\pi$ or five $2\pi$ pumping pulses. The decay function is close to exponential in all cases. The decay time increases by about $3\%$ when applying $\pi$ pumping pulses, but there is no improvement when the pulses correspond to $2\pi$ rotations. The traces have been normalized to account for the decay during the pulse sequence. The data were measured with $\DT=30\us$.
 (B) Individual traces of the averaged decay data shown in panel A, measured without (left panel), 5 pumping pulses with $2\pi$-rotations (middle panel), and 5 pumping pulses with $\pi$ rotations (right panel). 
 (C)  Standard deviation of the data in panel B, demonstrating a reduction in temporal shot-to-shot fluctuations in the presence of $\pi$ pumping pulses, but no improvement when the pumping pulses correspond to $2\pi$ rotations.
 }
\label{fig:2pi}
\end{figure}

\end{widetext}

\end{document}